%% file: main.tex
\documentclass[%
 reprint,
 superscriptaddress,
bibnotes,
 amsmath,amssymb,
 aps,
]{revtex4-2}
\usepackage{array}
\usepackage{amsmath}
\usepackage{multirow}
\usepackage{tabularx, makecell, booktabs} 
\usepackage{graphicx}
\usepackage{dcolumn}
\usepackage{bm}
\newcolumntype{L}{>{$}l<{$}} 
\usepackage{siunitx}
\usepackage{hyperref}

\begin{document}

\preprint{APS/123-QED}
\title{Evidence for a Spectral Break or Curvature in the Spectrum of Astrophysical Neutrinos from 5\,TeV to 10\,PeV}

\input{authorlist}

\collaboration{IceCube Collaboration}
\noaffiliation
%

\begin{abstract}

We report improved measurements of the all flavor astrophysical neutrino spectrum with IceCube by combining complementary neutrino samples in two independent analyses. Both analyses show evidence of a harder spectrum at energies below $\sim$30~TeV compared to higher energies where the spectrum is well characterized by a power law. The spectrum is better described by a log parabola or a broken power law, the latter being the preferred model. Both, however, reject a single power law over an energy range 5~TeV--10~PeV with a significance $>4\sigma$, providing new constraints on properties of cosmic neutrino sources. 

\end{abstract}
\maketitle


%
%
\textit{Introduction---} The IceCube South Pole Neutrino Observatory uses an array of 5160 optical modules (DOMs) deployed in a cubic kilometer of Antarctic glacial ice to detect the Cherenkov radiation of secondary particles produced in deep inelastic scattering (DIS) of neutrinos at TeV--PeV energies. 
High-energy astrophysical neutrinos, first observed by IceCube in 2013~\cite{IceCube:2013_HE,PRL_PeV,IceCube:2014_3HE}, arise from the interactions of  cosmic rays (CRs) with surrounding matter or photons in astrophysical sources and during their propagation through the Universe~\cite{Halzen:2022pez}. 
The measured all-sky diffuse flux of high-energy neutrinos represents the superposition of the neutrino emission from all sources in the observable Universe. 
An accurate characterization of the spectrum enables a better understanding of the dominant source populations and their relative contribution over the measured energy range~\citep{murase_hidden_2016,Murase:2019vdl,Fang:2022trf}. This helps elucidate the environments in which CRs are accelerated to produce neutrinos~\citep{anchordoqui2008high,Winter:2014tta}, and the relation between the sources of the neutrino emission and other probes of CR acceleration in the Universe~\cite{halzen2013pionic}. 
The spectrum may also contain signatures of new physics, e.g., from the annihilation of dark matter into neutrinos~\cite{DarkMatterAnnihilation}. Signatures of any of these processes can appear as features other than a simple power law in the neutrino spectrum. For example, a feature like a spectral break could indicate changes in the contributing source population, or help resolve the mechanisms for neutrino production, e.g. via p-$\gamma$ interactions.

IceCube has studied the cosmic neutrino flux with various detection morphologies \cite{IceCube:2013_HE,IceCube:2014_3HE,HESE7.5,IceCube:2015qii,IceCube:2016umi,IceCube:2018pgc,NorthernTracks,SBUCascades,MESE_2yr,ESTES,CNN_NuTau}.
Recorded events from neutrino interactions can be split into two main morphologies: 
cascades and tracks. Cascades, produced in charged-current (CC) electron and tau neutrino ($\nu_e$, $\nu_\tau$) DIS interactions, and neutral-current interactions of $\nu_e$, $\nu_\mu$, and $\nu_\tau$, can only be detected close to the instrumented volume. 
The energy transferred to the nucleus and the outgoing charged lepton (if present) is deposited in a particle shower over a few meters. 
The Cherenkov light yield is proportional to the deposited energy, with a typical energy resolution of $\sim \SI{8}{\percent}$ at \SI{100}{TeV}~\cite{2014JInst...9P3009A}. Tracks are generated when muon neutrinos ($\nu_\mu$) undergo CC DIS interactions, resulting in a Cherenkov light pattern along the trajectory of the secondary muon. A small fraction of them may also be created by the decay of a $\tau$ lepton created by CC DIS of $\nu_{\tau}$. The light yield is proportional to the muon energy loss above a few TeV in muon energy. 
As high-energy muons propagate for several kilometers in ice, the effective detection volume for tracks is much larger than the instrumented volume.
This yields high statistics with a good angular resolution ($0.3^{\circ}$ at 100~TeV) compared to cascades, but at the cost of energy resolution ($\sim65\%$ for the muon energy at \SI{100}{TeV}~\cite{2013NIMPA.703..190A}), as only a part of the Cherenkov light is deposited inside the detector volume.
 Starting events form a subset of both morphologies, where the neutrino-interaction vertex lies within the detector volume, depositing a majority of the initial hadronic energy in this volume. In particular, starting tracks are a subset of tracks with superior neutrino energy resolution ($ 
 \SI{26}{\percent}$ at \SI{100}{TeV}~\cite{ESTES}) due to the contained interaction vertex. The main backgrounds for studying astrophysical neutrinos are atmospheric neutrinos and muons produced by CR air showers, which are removed from event samples using dedicated selection techniques~\cite{IceCube:2013_HE,HESE7.5,NorthernTracks,SBUCascades,ESTES}. The atmospheric neutrino flux can be characterized as the ``conventional" flux, arising from the decay of pions and kaons created by CR interactions with atmospheric nuclei, and the ``prompt" flux, originating mainly from the decay of charmed hadrons in CR air showers. 

Here, we present two separate analyses, both leveraging the strengths of the cascade and track channels.
The first analysis, henceforth referred to as the ``Combined Fit" (CF), is based on the combination of existing data samples: tracks (focused on the Northern sky, where $\sim60\%$ events are through-going tracks) \cite{NorthernTracks}, and the improved all-sky contained cascades sample \cite{SBUCascades} with extended live time.
The second analysis expands the concept of high-energy starting events \cite{IceCube:2013_HE} to lower energies of a few TeV by selecting ``Medium Energy Starting Events" (MESE).
The MESE sample builds upon a prior analysis that used 2 years of IceCube data \cite{MESE_2yr}. Improved selection strategies, event reconstructions, and treatment of systematic uncertainties compared to previous studies have now been applied by the current MESE analysis.
The MESE events are classified as starting tracks and starting cascades to account for their different backgrounds and uncertainties.

\textit{Data samples---}
The CF (MESE) analysis uses data taken from May 2010 (May 2011) to May 2021 (June 2023), processed using a uniform calibration and filtering scheme \cite{2020JInst..15P6032A}. This is a significant improvement over a previous study combining IceCube track and cascade samples \cite{SBUCascades}, and is expanded on in a companion paper~\cite{PRD}.  Table~\ref{tab:event_numbers} shows the live time of each sample.
Simulated datasets based on the
in-ice light propagation models from \citep{2013NIMPA.711...73A,Chirkin:2013lpu} were used for both analyses, processed in exactly the same way as the experimental data.
The CF analysis characterizes the astrophysical flux using a combination of: 1)~a high-statistics muon track sample with good directionality and a high purity of muon neutrino events to constrain atmospheric neutrino flux and detector uncertainties; and 2)~a sample of contained cascades providing superior energy resolution, and a full-sky multiflavor acceptance. 
\begin{figure*}[t!]
    \begin{minipage}[b]{0.49\linewidth}
      \includegraphics[width=\linewidth]{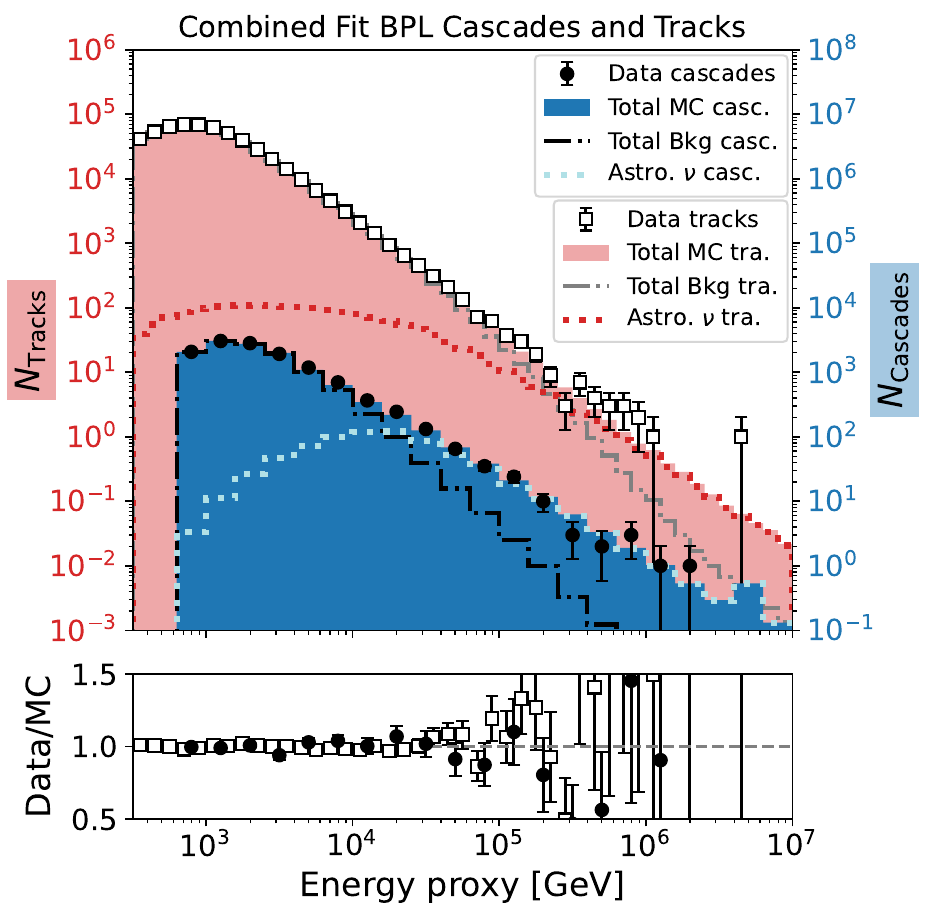}
    \end{minipage}
    \begin{minipage}[b]{0.49\linewidth}
      \includegraphics[width=\linewidth]{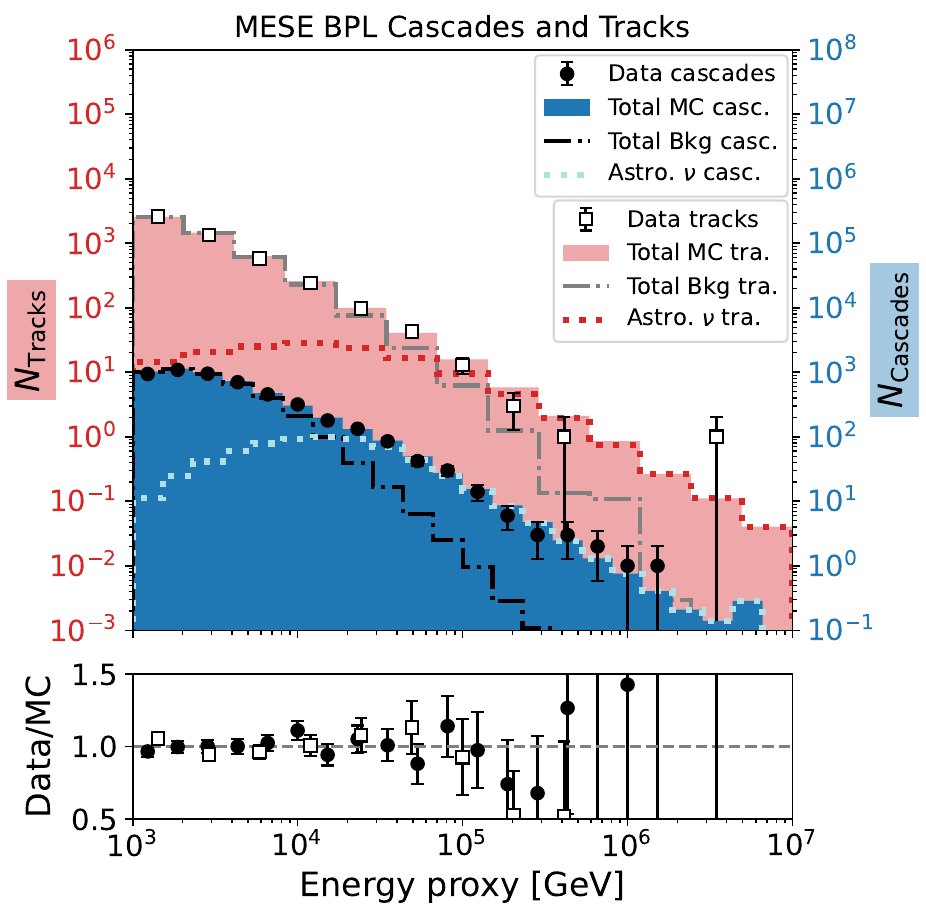}
    \end{minipage}
\caption{\textit{Energy distribution of cascade- and track-identified events used in the CF and MESE analyses:} contributions from astrophysical neutrinos, assuming the best-fit broken power-law spectrum, total background (atmospheric neutrinos and muons) along with the total contributions compared to data for CF on the left (8.5 years of tracks and 10 years of cascades) and MESE on the right (11.4 years). Note the different scales for tracks (left axis) and cascades (right axis). Here, energy proxy represents the reconstructed energy for the given event type which is different for cascades, MESE tracks, and CF tracks. Hence these energy proxies are not directly comparable.}
\label{fig:data_mc}
\vspace{-6mm}
\end{figure*}
\begin{table}[b]
\vspace{-5mm}
\setlength{\tabcolsep}{5pt}
\renewcommand{\arraystretch}{1.1}
\caption {Number of events in each data sample (reconstructed energy, $E_{\mathrm{reco}}$ $>$ \SI{1}{TeV}). The live time in years is shown in parentheses.}
\begin{tabular}{c|c|c|c}
 & MESE & CF & Overlap \\ \hline
Cascades & 4949 (11.4) & 10569 (10.5) & 2514 \\ \hline
Tracks & 4908 (11.4) & 231486 (8.5) & 1799
\end{tabular}
\label{tab:event_numbers}
\vspace{-6mm}
\end{table}
The MESE sample is composed of events that start inside the instrumented volume. This is achieved by rejecting events in veto regions that cover the boundaries of the detector. The size of the veto regions depends on the amount of light deposited in the detector by an event. The selected events are classified into tracks and cascades using a neural-network classifier~\cite{DNN_TheoGlauch}. Additional details regarding this refined event selection can be found in \cite{PRD}.
In both analyses, the cascades are most sensitive to astrophysical neutrinos with energies of $\mathcal{O}(\SI{10}{TeV})$. 
The proportion of atmospheric neutrinos in cascades is lower when compared to tracks due to the strong dominance of $\nu_{\mu}$ in the atmospheric spectrum at these energies \cite{riehn_hadronic_2018}.
In addition, the good energy resolution and a self-veto effect~\cite{PhysRevD.90.023009} also contribute to lowering the atmospheric neutrino background for cascades.
This self-veto effect, where muons accompanying the atmospheric neutrinos from the same down going cosmic-ray air shower trigger the veto, causes a suppression of the atmospheric-neutrino flux in the Southern sky. 
This introduces a strong angular dependence in the atmospheric neutrino flux in contrast to the extragalactic astrophysical neutrino flux which is expected to be isotropic. 
The samples used in the CF analysis and the MESE dataset are not independent, featuring a significant overlap especially for the cascade events. The number of events split by morphology in each sample and the corresponding overlap is presented in Table~\ref{tab:event_numbers}.

\textit{Method---}
The MESE and the CF datasets have been analyzed separately to measure the astrophysical spectrum of diffuse neutrinos.
Both analyses are based on a forward-folding binned likelihood technique, using the framework \textsc{NNMFit} \cite{PRD}. 
A simultaneous maximum-likelihood fit of the various components that contribute to the individual samples is performed, including the astrophysical flux, the conventional and prompt atmospheric neutrino background, and the atmospheric muon background. 
Systematic uncertainties are included in the form of nuisance parameters that affect the predicted observations. 
Cascades and tracks are binned individually in their respective reconstructed energy and zenith, as illustrated in Table~\ref{tab:Bins}.
\begin{table}[]
\vspace{-1mm}
\renewcommand{\arraystretch}{1.1}
\caption {2D Binning used in both analyses. $\theta$ is the reconstructed zenith angle. The binning is optimized for the sensitivity and the resolution of the observable in the respective morphology.}
\begin{tabular}{c|cc|cc}
             & \multicolumn{2}{c|}{Cascades}                & \multicolumn{2}{c}{Tracks}                   \\ \cline{2-5} 
Analysis     & \multicolumn{2}{c|}{$E_{\mathrm{reco}}$ (GeV) $|$ cos($\theta$)}
& \multicolumn{2}{c}{$E_{\mathrm{reco}}$ (GeV) $|$ cos($\theta$)} \\ \hline  
MESE         
& \multicolumn{1}{c|}{\begin{tabular}[c]{@{}c@{}}($10^3-10^7$)\\ 22 bins\end{tabular}} & \begin{tabular}[c]{@{}c@{}}(-1, 1)\\ 10 bins\end{tabular} & \multicolumn{1}{c|}{\begin{tabular}[c]{@{}c@{}}($10^3-10^7$)\\ 13 bins\end{tabular}} & \begin{tabular}[c]{@{}c@{}}(-1, 1)\\ 10 bins\end{tabular}    \\ \hline
\begin{tabular}[c]{@{}c@{}}Combined \\ Fit\end{tabular} & \multicolumn{1}{c|}{\begin{tabular}[c]{@{}c@{}}($4\times10^2-10^7$)\\ 22 bins\end{tabular}} & \begin{tabular}[c]{@{}c@{}}(-1, 1)\\ 3 bins\end{tabular}  & \multicolumn{1}{c|}{\begin{tabular}[c]{@{}c@{}}($10^{2.5}-10^7$)\\ 45 bins\end{tabular}} & \begin{tabular}[c]{@{}c@{}}(-1, 0.09)\\ 34 bins\end{tabular}
\end{tabular}
\label{tab:Bins}
\end{table}
Predictions for each flux component are obtained from the simulations mentioned above and are fitted to the data. The atmospheric neutrino fluxes were modeled from simulation and reweighted assuming flux estimates from the \textsc{MCEq} numerical framework ~\cite{fedynitch_calculation_2015,noauthor_mceq-projectmceq_2023}, assuming H4a~\cite{gaisser_spectrum_2012} as the primary CR composition model and Sibyll 2.3c~\cite{riehn_hadronic_2018} as the hadronic interaction model. 
The fit parameters include separate normalizations for the conventional and prompt fluxes. 
There are additional parameters for modifications to the primary cosmic ray spectrum, including a shift in its spectral index and a parameter that linearly interpolates the atmospheric neutrino spectrum predicted by the two primary CR composition models: H4a, which is dominated by protons at higher energies, and GST4~\cite{gaisser_cosmic_2013}, which assumes a more even mixture of particles at higher energies. 
We also account for variations in neutrino yields from $\pi/K$ decays in cosmic-ray showers~\cite{barr_uncertainties_2006}, as employed for the analysis of the large statistics through-going track sample~\cite{NorthernTracks}. We do not account for uncertainties in the prompt atmospheric flux with additional nuisance parameters, with more details provided in \cite{PRD}.
The atmospheric muon background is modeled by a kernel density estimator (KDE) generated from dedicated simulations~\cite{MuonGun} for MESE and the CF tracks, while we use the simulations directly for the CF cascades. The KDE compensates for the reduced availability of background simulation events in samples which reject a high proportion of muon events \cite{PRD}. The KDE is only used to model the background at the final level, after all selection cuts have been applied. Both analyses include an overall atmospheric muon normalization as a nuisance parameter in the fit. 
The modeling of the atmospheric self-veto uses the methodology of~\cite{arguelles_unified_2018}, but the two analyses use different parametrizations of the selection-dependent muon detection efficiency. 
Further details on the parametrizations in the fit can be found in the companion paper \cite{PRD}.

Various detector-related systematic uncertainties arise due to optical properties of the ice, such as the light absorption and  scattering coefficients, as well as anisotropic light propagation within the ice~\cite{IceModelAniso}, which affect the Cherenkov light patterns seen by the DOMs. 
In addition, the refreezing of the water surrounding the optical modules during deployment creates a column of ice filled with air bubbles, leading to a high local scattering coefficient. The effect of this \textit{hole-ice} on the photon angular acceptance of the DOMs is also modeled, as is a free parameter for their optical efficiency. 
These systematic effects are included in the maximum-likelihood fit via the SnowStorm method~\cite{aartsen_efficient_2019,ganster_combined_2022}. Based on simulated datasets that include systematic variations of the aforementioned ice and detector parameters, predictions of observed events are calculated from perturbations of these parameters around their nominal values.
The fit parameters within the CF analysis are largely unconstrained and the fit utilizes the large statistics of the tracks sample to self-consistently constrain its nuisance parameters. MESE, on the other hand, uses several priors on the nuisance parameters that modify the atmospheric flux and the detector response. These priors arise from previous IceCube calibration campaigns \cite{LEDCalibration}\cite{PRD}.

\begin{table*}[htbp]
\centering
\caption{Results for the spectral models tested in both analyses. The uncertainties are derived from 1D profile likelihood scans, assuming Wilks' theorem applies. We show the preference over the single power-law hypothesis in terms of the test statistic $\mathrm{TS}=-2 \Delta \mathrm{ln} \mathcal{L}$, where $\mathcal{L}$ is the likelihood value at the best fit for the given model. The form of the tested model for the total flux, $\Phi_{\nu+\bar{\nu}}$, is included in the table, and is measured in units of $\rm{10^{-18}/GeV/cm^2/s/sr}$. Here $\Lambda=\frac{E_{\nu}}{100\, \mathrm{TeV}}$,
$\gamma_{\mathrm{BPL}}=\left\{
\begin{array}{c}
       \gamma_1\, , (E_{\nu} < E_{\mathrm{break}})\\
       \gamma_2\, , (E_{\nu} > E_{\mathrm{break}})
\end{array}\right.$,
and $\phi_{0,\mathrm{broken}}=\phi_{0}\left\{
\begin{array}{c}
       \left(E_{\mathrm{break}}/100\,\mathrm{TeV}\right)^{-\gamma_1}\, , (E_{\mathrm{break}} > 100\,\mathrm{TeV})\\
       \left(E_{\mathrm{break}}/100\,\mathrm{TeV}\right)^{-\gamma_2}\, , (E_{\mathrm{break}} \leq 100\,\mathrm{TeV})
\end{array}\right.$.
All fluxes are normalized at \SI{100}{TeV}.
}

\label{tab:results}
\renewcommand{\arraystretch}{1.6}
\setlength\tabcolsep{0.8em}
\small

\begin{tabularx}{\linewidth}{c|c|c|c|c}
 & \multicolumn{4}{c}{Astrophysical model} \\ \cline{2-5} 
Analysis & \makecell[tc]{SPL\\$\left[
\phi_0 (\Lambda)^{-\gamma}\right]$ } & 
\makecell[tc]{SPE\\
$\left[
\phi_0 (\Lambda)^{-\gamma} e^{\frac{-E_{\nu}}{E_\mathrm{cutoff}}}\right]$
}
& 
\makecell[tc]{BPL\\$\left[
\phi_{0,\mathrm{broken}}(\frac{E_{\nu}}{E_{\mathrm{break}}})^{-\gamma_{\mathrm{BPL}}}\right]$
}
& 
\makecell[tc]{LP\\
$\left[
\phi_0 (\Lambda)^{-\alpha_\mathrm{LP}-\beta_\mathrm{LP}\log_{10}(\Lambda)
}\right]$
}
 \\ 
\cline{1-5} 
MESE & 
\begin{tabular}[t]{@{} l l @{}}
    $\phi_0$ & $= 2.13^{+0.18}_{-0.17}$ \\
    $\gamma$ & $= 2.55^{+0.04}_{-0.04}$
\end{tabular} & 
\begin{tabular}[t]{@{} l l @{}}
    $\phi_0$ & $= 3.98^{+1.14}_{-1.32}$ \\
    $\gamma$ & $= 2.16^{+0.23}_{-0.16}$ \\
    $\log_{10}(\frac{E_\mathrm{cutoff}}{\mathrm{GeV}})$ & $= 5.40^{+0.51}_{-0.23}$ \\
\end{tabular} & 
\begin{tabular}[t]{@{} l l @{}}
    $\phi_0$ & $= 2.28^{+0.22}_{-0.20}$ \\
    $\gamma_1$ & $= 1.72^{+0.26}_{-0.35}$ \\
    $\gamma_2$ & $= 2.84^{+0.11}_{-0.09}$ \\
    $\log_{10}(\frac{E_\mathrm{break}}{\mathrm{GeV}})$ & $= 4.52^{+0.11}_{-0.09}$ \\
\end{tabular} & 
\begin{tabular}[t]{@{} l l @{}}
    $\phi_0$ & $= 2.58^{+0.26}_{-0.26}$ \\
    $\alpha_\mathrm{LP}$ & $= 2.67^{+0.13}_{-0.06}$ \\
    $\beta_\mathrm{LP}$ & $= 0.36^{+0.10}_{-0.08}$ \\
\end{tabular} \\ \cline{2-5} 
 &
 & 
\begin{tabular}[t]{@{} l l @{}}
$-2 \Delta \mathrm{ln} \mathcal{L}$ & $= 1.8$
\end{tabular} & 
\begin{tabular}[t]{@{} l l @{}}
$-2 \Delta \mathrm{ln} \mathcal{L}$ & $= 27.3$
\end{tabular} & 
\begin{tabular}[t]{@{} l l @{}}
$-2 \Delta \mathrm{ln} \mathcal{L}$ & $= 18.8$
\end{tabular}\\ \cline{1-5} 
CF & 
\begin{tabular}[t]{@{} l l @{}}
    $\phi_0$ & $= 1.80^{+0.13}_{-0.16}$ \\
    $\gamma$ & $= 2.52^{+0.04}_{-0.04}$
\end{tabular} & 
\begin{tabular}[t]{@{} l l @{}}
    $\phi_0$ & $= 2.20^{+0.30}_{-0.25}$ \\
    $\gamma$ & $= 2.39^{+0.08}_{-0.08}$ \\
    $\log_{10}(\frac{E_\mathrm{cutoff}}{\mathrm{GeV}})$ & $= 6.15^{+0.37}_{-0.24}$ \\
\end{tabular} & 
\begin{tabular}[t]{@{} l l @{}}
    $\phi_0$ & $= 1.77^{+0.15}_{-0.11}$ \\
    $\gamma_1$ & $= 1.31^{+0.50}_{-1.21}$ \\
    $\gamma_2$ & $= 2.74^{+0.06}_{-0.07}$ \\
    $\log_{10}(\frac{E_\mathrm{break}}{\mathrm{GeV}})$ & $= 4.39^{+0.09}_{-0.08}$ \\
\end{tabular} & 
\begin{tabular}[t]{@{} l l @{}}
    $\phi_0$ & $= 2.13^{+0.16}_{-0.19}$ \\
    $\alpha_\mathrm{LP}$ & $= 2.57^{+0.06}_{-0.05}$ \\
    $\beta_\mathrm{LP}$ & $= 0.23^{+0.10}_{-0.07}$ \\
\end{tabular}\\ \cline{2-5}
 &
 & 
\begin{tabular}[t]{@{} l l @{}}
$-2 \Delta \mathrm{ln} \mathcal{L}$ & $= 7.5$
\end{tabular} & 
\begin{tabular}[t]{@{} l l @{}}
$-2 \Delta \mathrm{ln} \mathcal{L}$ & $= 24.7$
\end{tabular} & 
\begin{tabular}[t]{@{} l l @{}}
$-2 \Delta \mathrm{ln} \mathcal{L}$ & $= 16.4$
\end{tabular} 
\end{tabularx}
\end{table*}

\begin{figure}[t]
\includegraphics[width=0.97\linewidth]{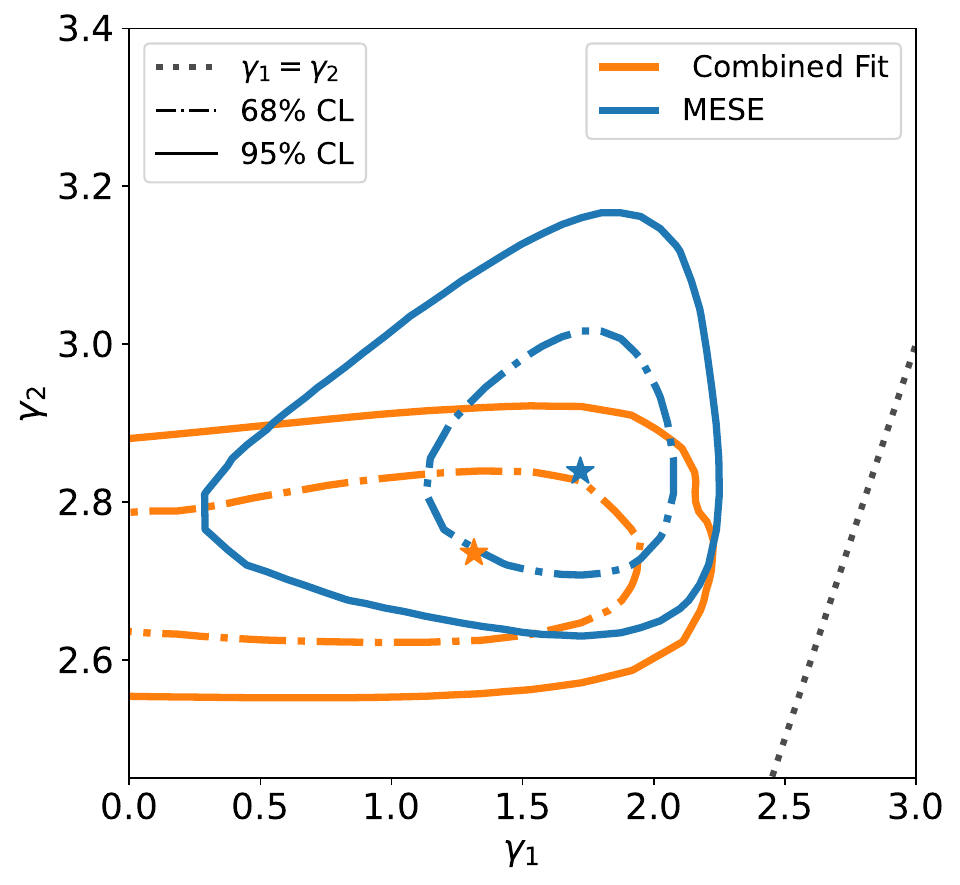}
\caption{\textit{Two-dimensional profile likelihood scan on the spectral indices of the broken power law:} the star markers indicate the best fit with $\gamma_1=1.72$ and $\gamma_2=2.84$ for MESE, and $\gamma_1=1.31$ and $\gamma_2=2.74$ for CF. Contours represent the 68\% and 95\% confidence regions based on Wilks’ theorem. The dotted line signifies the expectation of $\gamma_1=\gamma_2$, indicating the transition to a single power-law spectrum.}
\label{fig:bpl_2d}
\vspace{-7mm}
\end{figure}

\begin{figure}[t]
\includegraphics[width=1\linewidth]{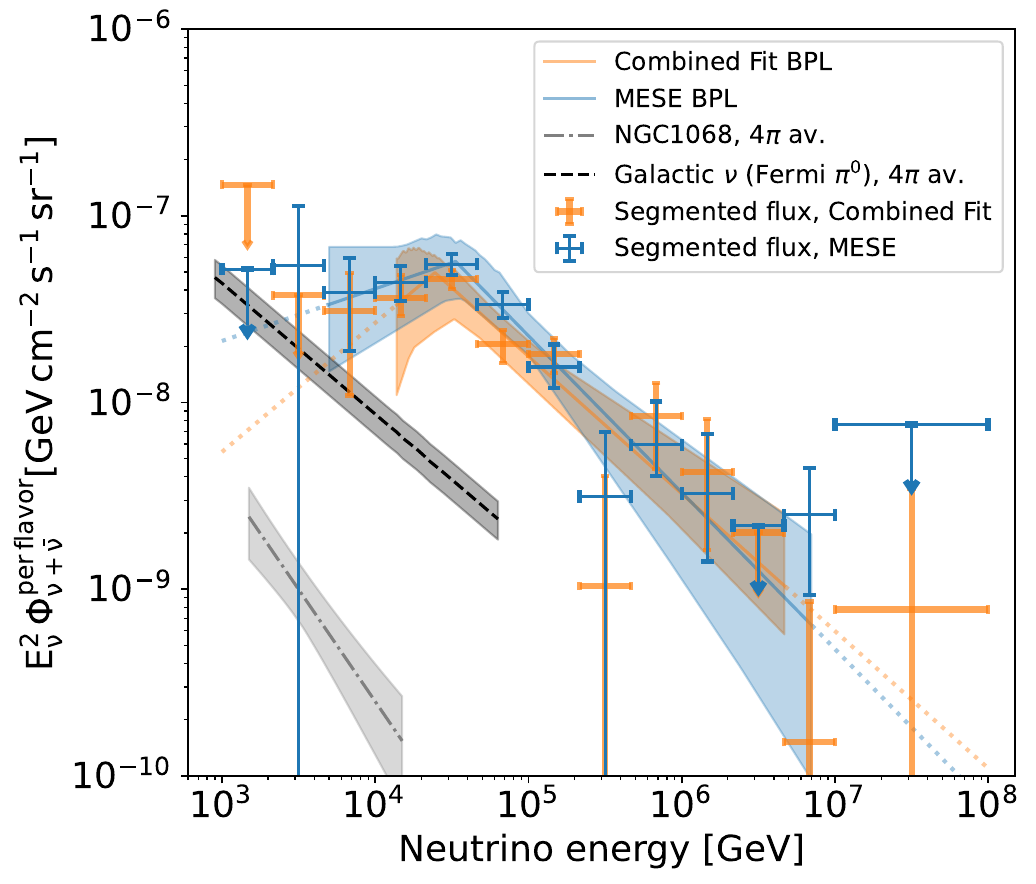}
\caption{\textit{Segmented flux:} the fit to the astrophysical flux normalization assuming an $E^{-2}$ spectrum is shown in each segment. Orange markers and shaded regions represent CF while MESE is shown in blue. The shaded regions show the 68\% uncertainties and sensitive energy ranges for each analysis obtained from profile likelihood scans of the four parameters of the BPL model. The flux from NGC1068~\cite{NGC1068} and the galactic plane~\cite{GalacticPlaneScience} measured with IceCube lie at much lower scales.}
\label{fig:neutrino_fluxes}
\vspace{-7mm}
\end{figure}

\textit{Results---}
We use the maximum-likelihood fit to test several hypotheses on the spectral shape of the astrophysical neutrino flux: 1)~a single power law (SPL); 2)~a single power law with an exponential cutoff (SPE), 3)~a log-parabolic (LP) flux;  and 4)~a broken power law (BPL) flux model. Both analyses find the likelihood to be maximum for the BPL model.  A list of the tested flux models is provided in Table~\ref{tab:results}, along with the functional forms, best fit parameters, and the difference in likelihood values when the model is compared to the SPL.
The SPE model test did not result in a significant improvement of the likelihood compared to the SPL model [$-2\Delta \mathrm{ln}\mathcal{L}=$~TS~$=\,7.5\,(1.8)$ for CF (MESE)]. 
The BPL model yields a noteworthy improvement of the TS by 24.4 (27.3) in the CF (MESE) analysis, corresponding to a significance of $4.4\,\sigma\,(4.7\,\sigma)$ over the SPL model based on Wilks' theorem \cite{Wilks}, with the TS distribution following a $\chi^2$ with two degrees of freedom (d.o.f.). A better TS is also noted for the LP model with TS = 16.4 (18.8) for CF (MESE) with a significance of $4.0\,\sigma\,(4.3\,\sigma)$.
Since the LP and BPL models have a different number of free parameters and are not nested, we calculate the preference of BPL over LP directly from the TS distribution obtained from pseudodata \cite{PRD}.
We find that the BPL is preferred over LP with a p value of 0.008 for the MESE analysis and a p value of 0.018 for the Combined Fit, when injecting the LP best fit and fitting with both models. 
A comparison of the agreement between data and Monte Carlo (MC) simulations assuming the best-fit parameters of the BPL is shown in Fig.~\ref{fig:data_mc}. 
Goodness of fit tests have been performed for each analysis and both morphologies, showing that they are statistically compatible.
We note a small deficit of data with respect to MC at a few hundred TeV. As it is not yet statistically significant, further investigation of this additional feature is required with more data. 
A comparison of the constraints on the low- and high-energy spectral indices obtained from the two analyses is shown in Fig.~\ref{fig:bpl_2d}. The figure illustrates the complementarity of the two analyses. The CF (MESE) analysis allows stronger constraints on the high-energy (low-energy) spectral index, as illustrated by their respective sensitive energy ranges (see Fig. \ref{fig:neutrino_fluxes}), \cite{PRD}. 
    
Finally, a segmented fit is performed, where the total neutrino flux, $E^2 \, \Phi_{\mathrm{astro}}^{\mathrm{total}} = \Sigma_{i} \, \phi_{i}\,\Theta(E-E_i)\,\Theta(E_{i+1}-E)$, is represented by the sum of fluxes $\phi_{i}$ in 13 independent energy bands $[E_{i},E_{i+1}]$ with a spectral index of 2 in each band, and $\Theta(x)$ represents the unit step function. It is important to note here that the fit is performed by independently fitting the contribution from each segment in true neutrino energy, rather than the reconstructed energy proxy.
This allows the characterization of energy-dependent features in the spectrum. The results are shown in Fig.~\ref{fig:neutrino_fluxes} and compared to the best-fit BPL model. 
There is generally good agreement between the segmented fit which has 14 d.o.f. and the measured BPL spectrum with 4 d.o.f., which is reflected in the small likelihood difference between them [TS = 6 (1.7) for CF (MESE)].

Various studies were performed on both samples to probe the robustness of the results by including additional nuisance parameters with appropriate priors. 
Uncertainties on the inelasticity in neutrino-nucleon DIS~\citep{NNSF,Eskola:2021nhw} were included by adding the mean inelasticity as a parameter to the fit using the description in~\cite{IceCube:2018pgc}. 
Data-driven parametrizations of the atmospheric neutrino spectrum~\cite{daemonflux} not included in the baseline fit were tested, as well as the impact of a neutrino flux from the galactic plane~\cite{GalacticPlaneScience} on the model's fit parameters. 
In addition, the data samples were split into data collected during the summer and winter months and fit separately. None of the tests lead to a significant change in the spectral parameters reported here. Details of the studies are discussed in \cite{PRD}.

\textit{Discussion and Conclusion---} Previous results from IceCube have hinted toward the existence of an excess ~\citep{MESE_2yr} or a break~\citep{SBUCascades} at $\sim \SI{30}{TeV} $ energy. 
This work has, for the first time, made a statistically significant observation of a break in the spectrum. The strength of both analyses reported here lies in the combination of complementary information from cascades and tracks. The larger statistics of the CF samples precisely constrains the high-energy spectral index and allows a self-consistent fit with minimal priors on systematic uncertainties. The MESE sample demonstrates superior sensitivity at energies below the break, leading to a better constraint on the low-energy spectral index \cite{PRD}.

We note that the contribution to the neutrino flux from the galactic plane~\cite{GalacticPlaneScience} and the brightest individual source NGC1068~\cite{NGC1068} to the measured total astrophysical neutrino spectrum reported here is minimal (see Fig. \ref{fig:neutrino_fluxes}). The neutrino spectrum of both NGC1068 and the Milky Way is softer than the diffuse neutrino spectrum below the break energy, indicating contributions from sources with a harder spectrum to the total extragalactic neutrino flux. 
An important consequence of our results is that the extragalactic neutrino flux at  $\mathcal{O}(10\,\mathrm{TeV})$ is lower compared to expectations from an SPL, favored by previous IceCube results. Various calculations hint toward the incompatibility of an SPL spectrum extrapolated to the \SIrange{1}{10}{TeV} energy range and the diffuse extragalactic gamma-ray spectrum~\cite{murase_hidden_2016,capanema_new_2020}. This is potentially alleviated by the spectrum reported here. Our results can also provide new constraints on the properties of extragalactic neutrino emitters (e.g.~\cite{fang_tev_2022}). Several theoretical models indeed predict a break or peak of the extragalactic diffuse emission in the TeV range~\cite{anchordoqui2017evidence}, which can now be refined.

\begin{acknowledgements}
The authors gratefully acknowledge the support from the following agencies and institutions:
USA {\textendash} U.S. National Science Foundation-Office of Polar Programs,
U.S. National Science Foundation-Physics Division,
U.S. National Science Foundation-EPSCoR,
U.S. National Science Foundation-Office of Advanced Cyberinfrastructure,
Wisconsin Alumni Research Foundation,
Center for High Throughput Computing (CHTC) at the University of Wisconsin{\textendash}Madison,
Open Science Grid (OSG),
Partnership to Advance Throughput Computing (PATh),
Advanced Cyberinfrastructure Coordination Ecosystem: Services {\&} Support (ACCESS),
Frontera and Ranch computing project at the Texas Advanced Computing Center,
U.S. Department of Energy-National Energy Research Scientific Computing Center,
Particle astrophysics research computing center at the University of Maryland,
Institute for Cyber-Enabled Research at Michigan State University,
Astroparticle physics computational facility at Marquette University,
NVIDIA Corporation,
and Google Cloud Platform;
Belgium {\textendash} Funds for Scientific Research (FRS-FNRS and FWO),
FWO Odysseus and Big Science programmes,
and Belgian Federal Science Policy Office (Belspo);
Germany {\textendash} Bundesministerium f{\"u}r Bildung und Forschung (BMBF),
Deutsche Forschungsgemeinschaft (DFG),
Helmholtz Alliance for Astroparticle Physics (HAP),
Initiative and Networking Fund of the Helmholtz Association,
Deutsches Elektronen Synchrotron (DESY),
and High Performance Computing cluster of the RWTH Aachen;
Sweden {\textendash} Swedish Research Council,
Swedish Polar Research Secretariat,
Swedish National Infrastructure for Computing (SNIC),
and Knut and Alice Wallenberg Foundation;
European Union {\textendash} EGI Advanced Computing for research;
Australia {\textendash} Australian Research Council;
Canada {\textendash} Natural Sciences and Engineering Research Council of Canada,
Calcul Qu{\'e}bec, Compute Ontario, Canada Foundation for Innovation, WestGrid, and Digital Research Alliance of Canada;
Denmark {\textendash} Villum Fonden, Carlsberg Foundation, and European Commission;
New Zealand {\textendash} Marsden Fund;
Japan {\textendash} Japan Society for Promotion of Science (JSPS)
and Institute for Global Prominent Research (IGPR) of Chiba University;
Korea {\textendash} National Research Foundation of Korea (NRF);
Switzerland {\textendash} Swiss National Science Foundation (SNSF).
\end{acknowledgements}
%
%


\textit{Data availability---} The data that support the findings of this article are openly available~\cite{DVN/ZBO52I_2026}
%
%
\nocite{*}
\bibliography{references}
\end{document}

%% file: authorlist.tex
\affiliation{III. Physikalisches Institut, RWTH Aachen University, D-52056 Aachen, Germany}
\affiliation{Department of Physics, University of Adelaide, Adelaide, 5005, Australia}
\affiliation{Dept. of Physics and Astronomy, University of Alaska Anchorage, 3211 Providence Dr., Anchorage, Alaska 99508, USA}
\affiliation{School of Physics and Center for Relativistic Astrophysics, Georgia Institute of Technology, Atlanta, Georgia 30332, USA}
\affiliation{Dept. of Physics, Southern University, Baton Rouge, Louisiana 70813, USA}
\affiliation{Dept. of Physics, University of California, Berkeley, California 94720, USA}
\affiliation{Lawrence Berkeley National Laboratory, Berkeley, California 94720, USA}
\affiliation{Institut f{\"u}r Physik, Humboldt-Universit{\"a}t zu Berlin, D-12489 Berlin, Germany}
\affiliation{Fakult{\"a}t f{\"u}r Physik {\&} Astronomie, Ruhr-Universit{\"a}t Bochum, D-44780 Bochum, Germany}
\affiliation{Universit{\'e} Libre de Bruxelles, Science Faculty CP230, B-1050 Brussels, Belgium}
\affiliation{Vrije Universiteit Brussel (VUB), Dienst ELEM, B-1050 Brussels, Belgium}
\affiliation{Dept. of Physics, Simon Fraser University, Burnaby, British Columbia V5A 1S6, Canada}
\affiliation{Department of Physics and Laboratory for Particle Physics and Cosmology, Harvard University, Cambridge, Massachusetts 02138, USA}
\affiliation{Dept. of Physics, Massachusetts Institute of Technology, Cambridge, Massachusetts 02139, USA}
\affiliation{Dept. of Physics and The International Center for Hadron Astrophysics, Chiba University, Chiba 263-8522, Japan}
\affiliation{Department of Physics, Loyola University Chicago, Chicago, IL 60660, USA}
\affiliation{Dept. of Physics and Astronomy, University of Canterbury, Private Bag 4800, Christchurch, New Zealand}
\affiliation{Dept. of Physics, University of Maryland, College Park, Maryland 20742, USA}
\affiliation{Dept. of Astronomy, Ohio State University, Columbus, Ohio 43210, USA}
\affiliation{Dept. of Physics and Center for Cosmology and Astro-Particle Physics, Ohio State University, Columbus, Ohio 43210, USA}
\affiliation{Niels Bohr Institute, University of Copenhagen, DK-2100 Copenhagen, Denmark}
\affiliation{Dept. of Physics, TU Dortmund University, D-44221 Dortmund, Germany}
\affiliation{Dept. of Physics and Astronomy, Michigan State University, East Lansing, Michigan 48824, USA}
\affiliation{Dept. of Physics, University of Alberta, Edmonton, Alberta, T6G 2E1, Canada}
\affiliation{Erlangen Centre for Astroparticle Physics, Friedrich-Alexander-Universit{\"a}t Erlangen-N{\"u}rnberg, D-91058 Erlangen, Germany}
\affiliation{Physik-department, Technische Universit{\"a}t M{\"u}nchen, D-85748 Garching, Germany}
\affiliation{D{\'e}partement de physique nucl{\'e}aire et corpusculaire, Universit{\'e} de Gen{\`e}ve, CH-1211 Gen{\`e}ve, Switzerland}
\affiliation{Dept. of Physics and Astronomy, University of Gent, B-9000 Gent, Belgium}
\affiliation{Dept. of Physics and Astronomy, University of California, Irvine, California 92697, USA}
\affiliation{Karlsruhe Institute of Technology, Institute for Astroparticle Physics, D-76021 Karlsruhe, Germany}
\affiliation{Karlsruhe Institute of Technology, Institute of Experimental Particle Physics, D-76021 Karlsruhe, Germany}
\affiliation{Dept. of Physics, Engineering Physics, and Astronomy, Queen's University, Kingston, Ontario K7L 3N6, Canada}
\affiliation{Department of Physics {\&} Astronomy, University of Nevada, Las Vegas, Nevada 89154, USA}
\affiliation{Nevada Center for Astrophysics, University of Nevada, Las Vegas, Nevada 89154, USA}
\affiliation{Dept. of Physics and Astronomy, University of Kansas, Lawrence, Kansas 66045, USA}
\affiliation{Centre for Cosmology, Particle Physics and Phenomenology - CP3, Universit{\'e} catholique de Louvain, Louvain-la-Neuve, Belgium}
\affiliation{Department of Physics, Mercer University, Macon, Georgia 31207-0001, USA}
\affiliation{Dept. of Astronomy, University of Wisconsin{\textemdash}Madison, Madison, Wisconsin 53706, USA}
\affiliation{Dept. of Physics and Wisconsin IceCube Particle Astrophysics Center, University of Wisconsin{\textemdash}Madison, Madison, Wisconsin 53706, USA}
\affiliation{Institute of Physics, University of Mainz, Staudinger Weg 7, D-55099 Mainz, Germany}
\affiliation{Department of Physics, Marquette University, Milwaukee, Wisconsin 53201, USA}
\affiliation{Institut f{\"u}r Kernphysik, Universit{\"a}t M{\"u}nster, D-48149 M{\"u}nster, Germany}
\affiliation{Bartol Research Institute and Dept. of Physics and Astronomy, University of Delaware, Newark, Delaware 19716, USA}
\affiliation{Dept. of Physics, Yale University, New Haven, Connecticut 06520, USA}
\affiliation{Columbia Astrophysics and Nevis Laboratories, Columbia University, New York, New York 10027, USA}
\affiliation{Dept. of Physics, University of Oxford, Parks Road, Oxford OX1 3PU, United Kingdom}
\affiliation{Dipartimento di Fisica e Astronomia Galileo Galilei, Universit{\`a} Degli Studi di Padova, I-35122 Padova PD, Italy}
\affiliation{Dept. of Physics, Drexel University, 3141 Chestnut Street, Philadelphia, Pennsylvania 19104, USA}
\affiliation{Physics Department, South Dakota School of Mines and Technology, Rapid City, South Dakota 57701, USA}
\affiliation{Dept. of Physics, University of Wisconsin, River Falls, Wisconsin 54022, USA}
\affiliation{Dept. of Physics and Astronomy, University of Rochester, Rochester, New York 14627, USA}
\affiliation{Department of Physics and Astronomy, University of Utah, Salt Lake City, Utah 84112, USA}
\affiliation{Dept. of Physics, Chung-Ang University, Seoul 06974, Republic of Korea}
\affiliation{Oskar Klein Centre and Dept. of Physics, Stockholm University, SE-10691 Stockholm, Sweden}
\affiliation{Dept. of Physics and Astronomy, Stony Brook University, Stony Brook, New York 11794-3800, USA}
\affiliation{Dept. of Physics, Sungkyunkwan University, Suwon 16419, Republic of Korea}
\affiliation{Institute of Physics, Academia Sinica, Taipei, 11529, Taiwan}
\affiliation{Dept. of Physics and Astronomy, University of Alabama, Tuscaloosa, Alabama 35487, USA}
\affiliation{Dept. of Astronomy and Astrophysics, Pennsylvania State University, University Park, Pennsylvania 16802, USA}
\affiliation{Dept. of Physics, Pennsylvania State University, University Park, Pennsylvania 16802, USA}
\affiliation{Dept. of Physics and Astronomy, Uppsala University, Box 516, SE-75120 Uppsala, Sweden}
\affiliation{Dept. of Physics, University of Wuppertal, D-42119 Wuppertal, Germany}
\affiliation{Deutsches Elektronen-Synchrotron DESY, Platanenallee 6, D-15738 Zeuthen, Germany}

\author{R. Abbasi}
\affiliation{Department of Physics, Loyola University Chicago, Chicago, IL 60660, USA}
\author{M. Ackermann}
\affiliation{Deutsches Elektronen-Synchrotron DESY, Platanenallee 6, D-15738 Zeuthen, Germany}
\author{J. Adams}
\affiliation{Dept. of Physics and Astronomy, University of Canterbury, Private Bag 4800, Christchurch, New Zealand}
\author{S. K. Agarwalla}
\thanks{also at Institute of Physics, Sachivalaya Marg, Sainik School Post, Bhubaneswar 751005, India}
\affiliation{Dept. of Physics and Wisconsin IceCube Particle Astrophysics Center, University of Wisconsin{\textemdash}Madison, Madison, Wisconsin 53706, USA}
\author{J. A. Aguilar}
\affiliation{Universit{\'e} Libre de Bruxelles, Science Faculty CP230, B-1050 Brussels, Belgium}
\author{M. Ahlers}
\affiliation{Niels Bohr Institute, University of Copenhagen, DK-2100 Copenhagen, Denmark}
\author{J.M. Alameddine}
\affiliation{Dept. of Physics, TU Dortmund University, D-44221 Dortmund, Germany}
\author{S. Ali}
\affiliation{Dept. of Physics and Astronomy, University of Kansas, Lawrence, Kansas 66045, USA}
\author{N. M. Amin}
\affiliation{Bartol Research Institute and Dept. of Physics and Astronomy, University of Delaware, Newark, Delaware 19716, USA}
\author{K. Andeen}
\affiliation{Department of Physics, Marquette University, Milwaukee, Wisconsin 53201, USA}
% ...
